\documentclass[cameraready]{jcmsi}%

\usepackage{graphicx}
\usepackage[fleqn]{amsmath}
\usepackage[psamsfonts]{amssymb}
\usepackage{mathtools}

\usepackage{txfonts}
\makeatletter%
\input{ot1txtt.fd}
\makeatother%

\usepackage[square,sort,comma,numbers]{natbib}

\usepackage{har2nat}

\usepackage{mathptmx}
\usepackage{enumerate}
\usepackage{lineno}
\usepackage{ntheorem}

\theoremstyle{plain}
\newtheorem{thm}{Theorem}
\newtheorem{lem}[thm]{Lemma}
\newtheorem{rmk}[thm]{Remark}
\newtheorem{defn}[thm]{Definition}
\newtheorem{assum}[thm]{Assumption}

\newtheorem{prob}[thm]{Problem}

\newtheorem*{pf}{Proof}

\DeclareMathOperator*{\diag}{diag}

\DeclareMathOperator*{\minimize}{\ \ \ \text{\upshape minimize}\ \ \ }
\DeclareMathOperator*{\subjectto}{\ \ \ \text{\upshape subject to}\ \ \ }

\usepackage{accents}
\newcommand{\ubar}[1]{\underaccent{\bar}{#1}}
\usepackage{color}
\definecolor{forestgreen}{rgb}{0.33,0.61,0.34}

\Vol{1}
\No{1}
\Month{1}
\Year{2020}

\title{Stability Optimization of Positive Semi-Markov Jump Linear Systems \\ via Convex Optimization}

\begin{document}

\AUTHOR{%
\author{Chengyan Zhao}{1},
\author{Masaki Ogura}{2},
\author{Kenji Sugimoto}{1}
}

\AFFILIATE{%
\affiliate{Graduate School of Science and Technology, Nara Institute of Science and Technology, Ikoma, Nara 630-0192, Japan}{1}
\affiliate{Graduate School of Information Science and Technology, Osaka University, Suita, Osaka 565-0871, Japan}{2}
}

\email{zhao.chengyan.za5@is.naist.jp,\\
m-ogura@ist.osaka-u.ac.jp,\\
kenji@is.naist.jp.}

\begin{abstract}%
	In this paper, we study the problem of optimizing the stability of positive semi-Markov jump linear systems. We specifically consider the problems of tuning the coefficients of the system matrices for maximizing the exponential decay rate of the system under a budget-constraint and minimizing the parameter tuning cost under the decay rate constraint. By using a result from the matrix theory on the log-log convexity of the spectral radius of nonnegative matrices, we show that the stability optimization problems are reduced to convex optimization problems under certain regularity conditions on the system matrices and the cost function. We illustrate the validity and effectiveness of the proposed results by using an example from the population biology.
\end{abstract}



\maketitle

\section{Introduction}

Markov jump linear systems~\cite{Costa2013} form an important class of stochastic switched dynamical systems and have applications in mobile robots~\cite{Bowling2006}, epidemic processes~\cite{Ogura2015c}, and networked control systems~\cite{Hespanha2007a}. Several important issues on Markov jump linear systems have been addressed in the literature including controllability and stabilizability~\cite{Ji1990}, robust optimal control~\cite{Costa1998a}, sampled-data control~\cite{Geromel2015}, and game theory~\cite{Mukaidani2017}. Furthermore, the class of systems includes the basic class of stochastic dynamical systems with an independent and identically distributed parameters~\cite{Ogura2012b,Hosoe2019}. Despite the aforementioned advances, modelling by a Markov jump linear system suffers from the limitation that the sojourn time of the systems must follow an exponential distribution. This restriction is not necessarily satisfied in practice; a typical example arises in modeling the occurrence of component failures in the context of fault tolerant control systems, where the probability density functions of failure rates are well-explained by Weibull distributions~\cite{Johnson1989}.

One natural way to overcome this limitation is to allow the sojourn times to follow non-exponential distributions, which results in a broader class of stochastic dynamical systems called \emph{semi-Markov jump linear systems}~\cite{Ogura2013f}. In this context, we can find in the literature several results toward the analysis and control of the class of systems. The authors in \cite{Schioler2014} have presented sufficient conditions for the moment stability of semi-Markov jump linear systems. Huang and Shi~\cite{Huang2012} derived linear matrix inequalities (LMIs) for the robust state-feedback control of semi-Markov jump linear systems. We can also find several LMI-based approaches for further advanced types of synthesis methodologies~\cite{Chen2016b,Jiang2018,Shen2018,Wei2017}. However, the design methodologies in the aforementioned references can be conservative because their derivation rely on approximations for avoiding the difficulty in dealing with non-exponential distributions. 

In this paper, we show that the problem of tuning the parameters of parametrized \emph{positive} semi-Markov jump linear systems can be efficiently and exactly solved without introducing any conservatism. We specifically show that, under the assumption that the parametrization is described by posynomial functions~\cite{Boyd2007}, the problems of finding the parameters maximizing the decay rate of the parametrized system and minimizing the parameter tuning cost can be transformed to convex optimization problems. The reduction to convex optimization problems is exact because we do not rely on any approximations that are employed in the aforementioned references. Instead of employing an approximation, in this paper we utilize the stability characterization of positive semi-Markov jump linear systems~\cite{Ogura2013f} as well as the log-log convexity result on the spectral radius of nonnegative matrices~\cite{kingman1961}. The theoretical result in this paper is illustrated by an example in the context of the population biology~\cite{Kussell2005}. 

This paper is organized as follows. In Section~\ref{sec2}, we formulate the stabilization problem of positive semi-Markov jump linear systems and state the  main result. The derivation of the main result is presented in Section~\ref{sec3}. In Section~\ref{sec4}, we illustrate the validity and effectiveness of the result with numerical simulations. 

\subsection*{Notations} Let $(\Omega, M, P)$ be a probability space. The expected value of a random variable $X$ on $\Omega$ is denoted by $E[X]$. Let $\mathbb{R}$, $\mathbb{R_+}$, and $\mathbb{R}_{++}$ denote the set of real, nonnegative, and positive numbers, respectively. Let $\mathbb{R}^{n\times n}$ denote the set of $n \times n$ real matrices. The identity matrix of order~$n$ is denoted by $I_n$. We let $x\geq 0$ be a non-negative vector, if the entries of $x$ are all nonnegative. We say that a square matrix is Metzler if the off-diagonal entries of the matrix~$A$ are nonnegative. We denote the spectral radius of~$A$ by $\rho(A)$. We define the entrywise logarithm of a vector $v \in \mathbb{R}^n_{++}$ by $\log[v]=[\log v_1, \dots ,\log v_n]^\top$. The entrywise exponential operation $\textup{exp}[\cdot]$ is defined in the same manner.

\section{Main results}\label{sec2}

Let us consider a parameterized family of switched linear systems of the form  
\begin{equation}\label{eq:sigmaTheta}
\Sigma_\theta: \frac{dx}{dt}=A_{\sigma(t)}(\theta)x(t),\quad x(0) = x_0 \in \mathbb{R}^n, 
\end{equation}
where $x(t) \in \mathbb{R}^{n}$ is the state vector, $\sigma$ is a piecewise-constant function taking values in the set~$\{1, \dots, N \}$, and $A_1(\theta)$, \dots, $A_N(\theta)\in \mathbb{R}^{n \times n}$ are matrices parametrized by the parameter~$\theta$ belonging to a set~$\Theta \subset \mathbb{R}^\ell$. 

In this paper, we assume that each subsystem has a positivity (see, e.g.,~\cite{Farina2000,Ebihara2017a}). We also assume that $\sigma$ is a semi-Markov process~\cite{Janssen2006}; i.e., we assume that the evolution of $\sigma$ is governed by the following probabilities: 
\begin{equation*}
\textup{Pr}\{\sigma(t+h)=j \mid \sigma(t)=i\}=
\left\{ 
\begin{aligned}
&\lambda_{ij}(h)h+o(h),  \quad~~~  \mbox{if } j\neq i,\\
&1+\lambda_{ii}(h)h+o(h),~ \mbox{if } j = i,
\end{aligned}
\right.
\end{equation*}
where $\lambda_{ij}(h)$ represents a time-varying transition rate from mode $i$ to mode $j$, $\lambda_{ii}(h)=-\sum_{j=1,i\neq j}^N\lambda_{ij}(h)$, and $o(h)$ is little-$o$ notation defined by $\lim_{h \to 0}o(h)/h=0$. The above assumptions are summarized into the following definition. 

\begin{defn}[\cite{Ogura2013f}]\label{defn:system}
Let $\theta \in \Theta$. We say that the system~$\Sigma_\theta$ is a \emph{positive semi-Markov jump linear system} if the initial state~$x_0$ is nonnegative, the matrices~$A_1(\theta)$, \dots,~$A_N(\theta)$ are Metzler, and $\sigma$ is a semi-Markov process taking values in $\{1, \dotsc, N\}$. 
\end{defn}

For $t\geq 0$ and $x_0 \in \mathbb{R}_+^n$, we let $x(t; x_0)$ denote the trajectory of the system~$\Sigma_\theta$ at time $t$ and with the initial condition $x(0) = x_0$. This paper is concerned with the stability property of the system~$\Sigma_\theta$ given as follows:

\begin{defn}[\cite{Feng1992,Ogura2013f}]
We say that $\Sigma_\theta$ is \emph{exponentially mean stable} if there exist  $\alpha>0$ and $\beta>0$ such that, for every $x_0$ and $\sigma(0)$,
\begin{equation*}
 E[\|x(t; x_0)\|] \leq \alpha e^{-\beta t}\|x(0)\|.
\end{equation*}
If $\Sigma_\theta$ is mean stable, then the \emph{exponential decay rate} of the system~$\Sigma_\theta$ is defined by
\begin{equation*}
\gamma_\theta =- \sup_{x_0\in\mathbb{R}^n_+} \limsup_{t\to\infty}\frac{\log E[\|x(t; x_0)\|]}{t}. 
\end{equation*}
\end{defn}

In this paper, we consider a budget-constrained stability optimization problem described as follows. Consider the situation where a limited amount of resource available is given for tuning the parameter~$\theta$ to improve the stability of the system~$\Sigma_\theta$. We let a real function~$C$ denote the cost for achieving a specific parameter~$\theta$. In this context, we formulate our stability optimization problem as follows:

\begin{prob}[Budget-constrained stabilization]\label{prb:}
Let a real number~$\bar C$ be given. Find the parameter~$\theta \in \Theta$ such that the exponential decay rate~$\gamma_\theta$ is maximized, while the budget constraint
\begin{equation*}
	C(\theta) \leq \bar C
\end{equation*}
is satisfied. 
\end{prob}

In the budget-constrained
optimization problem, we need to distribute the constrained parameter cost to $A_i(\theta)$ to obtain the maximized decay rate. However, there is another situation where the optimization object is minimizing the parameter tuning
cost $C(\theta)$ while satisfying the performance constraint (decay rate). From this perspective, we formulate an alternative optimization problem as follows:

\begin{prob}[Performance-constrained stabilization]\label{prb:performance}
	Let a positive number~$\bar \gamma$ be given. Find the parameter~$\theta \in \Theta$ such that the parameter tuning cost $C(\theta)$~is minimized, while the performance constraint
	\begin{equation*}
		\gamma_
		\theta \geq \bar \gamma
	\end{equation*}
	is satisfied. 
\end{prob}

In our main results, we show that Problems~\ref{prb:} and~\ref{prb:performance} reduce to convex optimization problems. In order to state the main results, we place certain regularity assumptions on the system matrices~$A_i(\theta)$ and the cost function~$C(\theta)$. For this purpose, we introduce the class of functions called monomials and posynomials~\cite{Boyd2007}. We say that a function $F\colon \mathbb{R}_{++}^n \to \mathbb{R}_{++}$ is a \emph{monomial} if there exist $c > 0$ and real numbers $a_1$, \dots,~$a_n$ such that 
\begin{equation*}
	F(v)=cv_1^{a_1}v_2^{a_2}\dotsm v_n^{a_n}	
\end{equation*}
Then, we say that a function $F$ is a \emph{posynomial} if $F$ is a sum of finite number of monomials. 

The following mild and reasonable assumption is necessary for ensuring Problems~\ref{prb:} and~\ref{prb:performance} reduce to  convex optimization problems.

\begin{assum}\label{asm:}
The following conditions hold true:
\begin{enumerate}
\item \label{asm:metzler} For each $k = 1$, \dots, $N$, there exists an $n\times n$ Metzler matrix $M_k$ such that each entry of the matrix
\begin{equation*}
	\bar{A}_k(\theta)=A_k(\theta) - M_k
\end{equation*}
is either a posynomial in $\theta$ or zero. 

\item $C(\theta)$ is a posynomial in $\theta$. 

\item \label{asm:posyconstraint} There exist posynomials  $\phi_1(\theta)$, \dots, $\phi_m(\theta)$ and positive constants $\bar \phi_1$, \dots, $\bar \phi_m$ such that

	\begin{equation*}
\Theta = \{\theta\in\mathbb{R}^\ell\colon \phi_1(\theta)\leq \bar \phi_1, \dotsc, \phi_m(\theta)\leq \bar \phi_m \}. 
\end{equation*}

\item \label{asm:sojourn}The sojourn times of the semi-Markov process $\sigma$ are uniformly bounded, i.e., there exists $T>0$ such that the sojourn times are less than or equal to $T$ with probability one. 
\end{enumerate}
\end{assum}

\begin{rmk}\upshape
Major examples of positive linear time-invariant systems satisfying Assumption~\ref{asm:} include networked epidemic processes~\cite{Preciado2014}, population dynamics~\cite{Kussell2005}, and dynamical buffer networks~\cite{Rantzer2018} (for further discussions, see \cite{Ogura2019c}). For example, in the containment problem for networked epidemic processes~\cite{Preciado2014}, the parameter~$\theta$ corresponds to the infection and recovery rates of the nodes, while the cost function~$C(\theta)$ would indicate the cost for medical resources to tune the rates. 
\end{rmk}

Also, we introduce the following notations to state the main results of this paper. Let $\sigma_d$ be the embedded Markov chain of $\sigma$ (see, e.g., \cite{Janssen2006}). For $i, j\in\{1, \dotsc, N\}$, let $p_{ij}$ denote the transition probability of $\sigma_d$, i.e., let $p_{ij}$ denote the probability that the discrete-time Markov chain~$\sigma_d$ transitions into state~$j$ from state~$i$ in one time step. Also, let $h_{ij}$ denote the random variable representing the sojourn time of $\sigma$ at the mode $j$ after jumping from the mode~$i$, and $f_{ij}(h_{ij})$ denote the corresponding probability density function. 


\begin{thm}\label{thm:}
Let $\bar C> 0$ be given. For each $\theta \in \Theta$ and $g > 0$, define the matrix~$\mathcal A(\theta, g) \in \mathbb{R}^{{(nN)} \times {(nN)}}$ as the block matrix whose $(i,j)$-block is defined by
\begin{equation}\label{thm:eq:calA}
[\mathcal A(\theta, g)]_{ij}
=p_{ji} \int_{0}^{T}e^{(A_j(\theta)+g I)\tau}f_{ji}(\tau) \, d\tau \in \mathbb{R}^{n\times n}. 
\end{equation}
Define the set
\begin{equation}\label{thm:set}
	\log[\Theta] = \{ \log[\theta]\colon \theta \in \Theta \} \subset \mathbb{R}^\ell.
\end{equation}
Assume that $u=u^\star$ and $v = v^\star$ solve the following optimization problem: 
\begin{equation}\label{eq:convexOpt}
\begin{aligned}
 \minimize_{\mathclap{u\in\log[\Theta],\,v\in \mathbb{R}}}\ \ & -v
\\
\subjectto\ \ & \log\rho(\mathcal A(\exp[u], e^v)) \leq 0, 
\\
& \log C(\exp[u])\leq \log \bar C, 
\\
& \log \phi_i(\exp[u])\leq \log \bar \phi_i, \quad i=1, \dotsc, \ell. 
\end{aligned}
\end{equation}
Then, 
\begin{equation*}
\hspace{0.2cm}	\theta = \exp[u^\star]
\end{equation*}
solves Problem~\ref{prb:} and attains the exponential decay rate $e^{v^\star}$. Furthermore, the optimization problem~\eqref{eq:convexOpt} is convex. 
\end{thm}

We can also show that Problem~\ref{prb:performance} can be solved by the following convex optimization problem.

\begin{thm}\label{thm:performance:}
Assume that $u=u^\star$ solves the following optimization problem: 
\begin{equation}\label{thm:performance:eq:convexOptB}
\begin{aligned}
\minimize_{\mathclap{u\in\log[\Theta],\,v\in \mathbb{R}}}\ \ & \log C(\exp[u])
\\
\subjectto\ \ & \log\rho(\mathcal A(\exp[u], e^v)) \leq - \log \bar{\gamma}, 
\\
& \log \phi_i(\exp[u])\leq \log \bar \phi_i,\notag
\\
&\hspace{2cm}i=1, \dotsc, \ell. 
\end{aligned}
\end{equation}
Then, 
\begin{equation*}
\hspace{0.2cm}	\theta = \exp[u^\star]
\end{equation*}
solves Problem~\ref{prb:performance} and attains the minimized cost $C(\exp[u^\star])$. Furthermore, the optimization problem~(\ref{thm:performance:eq:convexOptB}) is convex. 
\end{thm}

\section{Proof}\label{sec3}

In this section, we give the proof of the main results. Because the proof of Theorem~\ref{thm:performance:} is similar to that of Theorem~\ref{thm:}, we only present the proof of Theorem~\ref{thm:}. We first prepare a few lemmas for the proof. The first lemma gives a characterization of the exponential decay rate of the system~$\Sigma_\theta$ in terms of the spectral radius of the matrix~$\mathcal A(\theta, g)$ defined in the theorem. 

\begin{lem}\label{lem:first}
Let $\theta\in\Theta$ and $g > 0$ be arbitrary. The following statements are equivalent: 
\begin{itemize}
	\item The exponential decay rate of $\Sigma_\theta$ satisfies $\gamma_\theta > g$. 
	\item $\rho(\mathcal A(\theta, g)) < 1$. 
\end{itemize}
\end{lem}

\begin{pf} \upshape
Assume $\gamma_\theta > g$. Then, the positive semi-Markov jump linear system 
\begin{equation*}
	\frac{dx}{dt} =(A_{\sigma(t)}(\theta) + g I)x(t)	
\end{equation*}
is exponentially mean stable. Therefore, by Theorem~2.5 in \cite{Ogura2013f}, the matrix $\mathcal A(\theta, g)$ has a spectral radius less than one, as desired. The proof of the opposite direction can be proved in the same manner and, therefore, is omitted.

We then recall the following celebrated result by \cite{kingman1961}. We say that an $\mathbb{R}_{++}$-valued function $f(x)$ is superconvex if $\log f(x)$ is convex.

\begin{lem}[\cite{kingman1961}]\label{lem:kingman}
Let $A\colon \mathbb{R}^\ell \to \mathbb{R}^{n\times n}_{++}$ be a function. Assume that each entry of $A$ is either a superconvex function or the zero function. Then, the mapping $\mathbb{R}^\ell \to \mathbb{R}_{++} \colon x \mapsto  \rho(A(x))$ is superconvex.
\end{lem}

We finally state the following lemma concerning the superconvexity of posynomials. 

\begin{lem}[{\cite{Boyd2007}}]\label{lem:boyd}
Let $f\colon \mathbb{R}^n_{++}\to \mathbb{R}_{++}$ be a posynomial. Then, the mapping
$\mathbb{R}^{n} \rightarrow \mathbb{R}_{++}:u \mapsto f(\textup{exp}[u	])$
is superconvex.
\end{lem}

Let us now prove Theorem~\ref{thm:}. 





Lemma~\ref{lem:first} shows that the solution of Problem~\ref{prb:} is given by the following optimization problem: 
\begin{equation}\label{eq:opt:original}
\begin{aligned}
\minimize_{\mathclap{\theta \in \Theta,\, g> 0}}\ \ & -g
\\
\subjectto\ \ & \rho(\mathcal A(\theta, g)) \leq 1, 
\\
& C(\theta)\leq \bar C, 
\\
& \phi_i(\theta) \leq \bar \phi_i, \ i=1, \dotsc, \ell.
\end{aligned}
\end{equation}
Performing the variable transformations
\begin{equation*}
 u = \log[\theta],\quad v = \log g
\end{equation*}
as well as taking logarithms in the objective functions and constraints, we can equivalently reduce \eqref{eq:opt:original} into the optimization problem in Theorem \ref{thm:}. Therefore, to complete the proof of theorem, we need to show the convexity of the optimization problem in Theorem \ref{thm:}. The convexity of the constraints in Theorem \ref{thm:} is a direct consequence of the superconvexity of posynomials stated in Lemma~\ref{lem:boyd}. In the remaining of this section, we shall show the convexity of the mapping
\begin{equation}\label{eq:mapping}
\log[\Theta]\times \mathbb{R}\to \mathbb{R}\colon (u, v)\mapsto \log\rho(\mathcal A(\exp[u], e^v)). 
\end{equation}
For each $k=1$, \dots, $N$, we define the matrix function
\begin{equation*}
	\tilde A_k(\theta) = A_k(\theta) - M_k+ gI.
\end{equation*}
Then, equation~(\ref{thm:eq:calA}) shows that
\begin{equation*}
	[\mathcal A(\theta, g)]_{ij} = {p_{ji}} \int_0^T e^{(\tilde A_j(\theta)+M_j)\tau }f_{ji}(\tau )\, d\tau,
\end{equation*}
where $f_{ji}$ denotes the probability density function of the sojourn time~$h_{ji}$. This equation and the Lie-product formula 
\begin{equation*}
	e^{A+B}=\lim_{K \to \infty}(e^{A/K}e^{B/K})^K	
\end{equation*}
for square matrices $A$ and $B$ (see, e.g., \cite{Cohen1981}) yield that
\begin{equation*}\label{eq:weobtaiin}
\begin{aligned}
[\mathcal A(\theta, g)]_{ij}
&=
{p_{ji}} \int_0^T\lim_{K\to\infty}\left(
e^{\frac{\tau \tilde A_j(\theta, g)}{K}} e^{\frac{\tau M_j}{K}}
\right)^Kf_{ji}(\tau) \, d\tau
\\
&=
\lim_{K, L\to\infty} \Gamma_{ij}^{(K, L)}(\theta, g) 
\end{aligned}
\end{equation*}
where, for positive integers~$K$ and~$L$, the $n\times n$ matrix~$\Gamma_{ij}^{(K, L)}(\theta, g)$ is defined by 
\begin{equation*}\label{eq:GammaKL}
\Gamma_{ij}^{(K, L)}(\theta, g) = {p_{ji}} \sum_{\ell=1}^L \frac{T}{L} \left(
e^{\frac{\ell T \tilde A_j(\theta, g)}{KL}} e^{\frac{\ell T M_j}{KL}}
\right)^K f_{ji}(\ell T/L).
\end{equation*}
Therefore, if we define 
\begin{equation*}\label{eq:GammaKLN}
\begin{multlined}
\Gamma_{ij}^{(K, L, M)}(\theta, g) =  \\ {p_{ji}}\sum_{\ell=1}^L  \frac{T}{L}
\left( \sum_{m=0}^M \frac{1}{m!} \left(\frac{\ell T \tilde A_j(\theta, g)}{KL}\right)^m\!\!  e^{\frac{\ell T M_j}{KL}}
\right)^K \!\!f_{ji}\!\left(\frac{\ell T}{L}\right)
\end{multlined}
\end{equation*}
then, by the definition of matrix exponentials, we obtain the following expression: 
\begin{equation}\label{eq:calAlimLKM}
[\mathcal A(\theta, g)]_{ij} = \lim_{K, L, M\to\infty}\Gamma_{ij}^{(K, L, M)}(\theta, g).
\end{equation}

Let us show that each entry of the matrix $\Gamma_{ij}^{(K, L, M)}$ is either a posynomial in $\theta$ and $g$ or zero. Since the matrix $M_j$ is assumed to be Metzler (Assumption~\ref{asm:}.\ref{asm:metzler}), the matrix~$e^{{\ell T M_j}/{KL}}$ is nonnegative for all $K$ and~$L$. Also, each entry of the matrix~$\tilde A(\theta, g)$ is either a posynomial or zero by Assumption~\ref{asm:}.\ref{asm:metzler}. Since the set of posynomials is closed under additions and multiplications, each entry of the matrix power $({\ell T \tilde A_j(\theta, g)}/{KL})^m$ is either a posynomial of $\theta$ and $g$ or zero as well. From the above observation, we conclude that each entry of the matrix~$\Gamma^{(K, L, M)}(\theta, g)$ is a posynomial with the variables~$\theta$ and $g$ or zero.

We are now ready to complete the proof of the theorem. Define the $(nN)\times (nN)$ matrix $\mathcal A^{(K, L, M)}(\theta, g)$ as the block matrix whose $(i,j)$-block equals $\Gamma_{ij}^{(K, L, M)}(\theta, g)$ for all $i, j \in \{1, \dotsc, N\}$. Then, by Lemmas~\ref{lem:kingman} and~\ref{lem:boyd}, the mapping 
\begin{equation*}
(u, v)\mapsto \rho(\mathcal A^{(K, L, M)}(\exp[u], e^v))
\end{equation*}
is superconvex. Since \eqref{eq:calAlimLKM} shows that the mapping $\mathcal A$ is a point-wise limit of the mapping~$\mathcal A^{(K, L, M)}$, taking a limit preserves superconvexity, and the spectral radius operator~$\rho(\cdot)$ is continuous, we obtain the convexity of the mapping~(\ref{eq:mapping}). This completes the proof of convexity of the optimization problem~(\ref{eq:convexOpt}), as desired.
\end{pf}

\begin{rmk}\upshape
From the proof of Theorem~\ref{thm:}, we see that   Problem~\ref{prb:} can be formulated as the problem~(\ref{eq:opt:original}) even without Assumption~\ref{asm:}. 
 Assumption~\ref{asm:} then allows us to reduce the optimization problem (6) into a convex optimization problem. 
\end{rmk}


\section{Numerical example}\label{sec4}

In this section, we illustrate the validity and effectiveness of the main results with an example in the context of population biology~\cite{Kussell2005}. Many biological populations are exposed to environmental fluctuations, from daily regular cycles of light and temperature to irregular fluctuations of nutrients and pH levels. To survive through the fluctuating environment, many biological populations employ a protection mechanism called \emph{bet-hedging} to increase robustness against the fluctuations of environment. In brief, bet-hedging means that the biological populations exhibit several phenotypes that have different growth rates among the possible environments.

\begin{figure}[tb]
	\centering
	\includegraphics[scale=0.27]{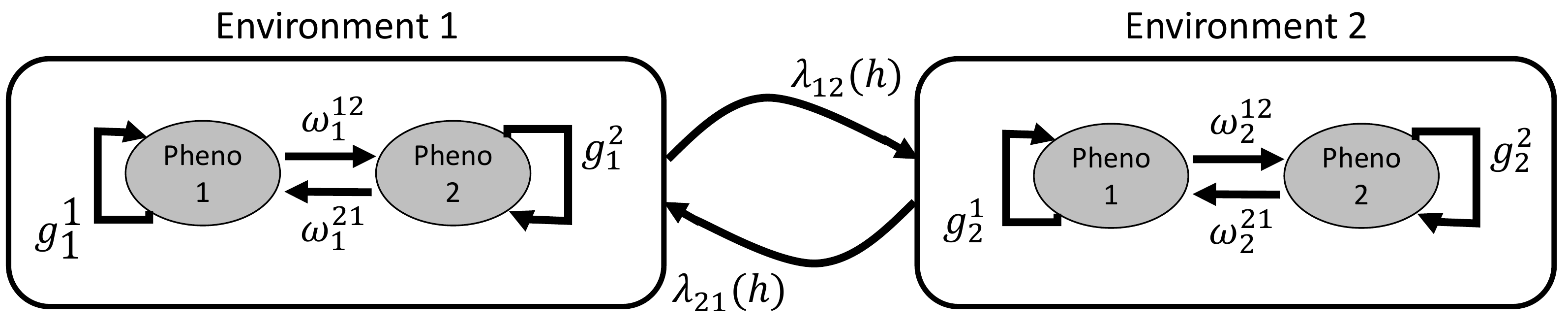}    
	\caption{Bet-hedging population ($n=2$ and $N=2$).}  
	\label{fig:bet} 
\end{figure}	

For the bet-hedging population model, we consider a biological community with $n$ phenotypes living in a randomly fluctuating environment with $N$ possible environmental types. Let $g^i_k$ denote the growth rate of phenotype $i$ under environment~$k$. In the bet-hedging population, individuals may switch their phenotype at any time but stochastically. We let $\omega^{ji}_k$ denote the instantaneous rate at which an individual having phenotype $j$ switches its phenotype to $i$ under environment $k$. Let $x_i(t)$ denote the number of individuals having phenotype $i$ at time~$t$ and $\sigma(t)$ denote the environment type at time $t$. Then, the dynamics of population in phenotype $i$ can be expressed by the following differential equation~\cite{Kussell2005} 
\begin{equation}\label{eq:dx_k/dt:withoutdelay}
\Sigma: \frac{dx_i}{dt}=g^i_{\sigma(t)}x_i(t)
+\sum_{j \neq i}^n \omega^{ji}_{\sigma(t)}x_{j}(t), 
\end{equation}
where $\omega^{ii}_k=-\sum_{j=1, j\neq i}^{n} \omega^{ij}_k$. The fluctuation is governed by a semi-Markov process $\sigma(t) \in \{1,\dotsc, N\}$ as mentioned in Definition \ref{defn:system}. Fig. \ref{fig:bet} shows a schematic picture of this model for $n=2$ and $N=2$, i.e., individuals present two types of phenotypes in two environments. 

Let us consider the problem of driving the entire population into extinction through biological intervention. Assume that $L$ different types of antibiotics are available for suppressing growth rates. Let $c_\ell(\alpha_\ell)$ ($\ell \in \{1,\dots,L\}$) denote the cost for dosing $\alpha_\ell$ unit of the $\ell$th antibiotics, which is assumed to reduce the growth rate of the $i$th phenotype population by $\Delta_\ell g^{i}(\alpha_\ell)$ independent of the current environment types. In this situation, we can reduce the growth rate of the $i$th phenotype population to $g^i_{\sigma(t)}-\sum_{\ell=1}^{L}\Delta_{\ell} g^i(\alpha_\ell)$ with the associated total cost $C(\alpha) = \sum_{\ell=1}^L c_\ell(\alpha_\ell)$. The resulting population dynamics admits the representation 
\begin{equation*}
\Sigma': \frac{dx_i}{dt}=\left(g^i_{\sigma(t)}-\sum_{\ell=1}^{L}\Delta_{\ell} g^i(\alpha_\ell)\right)x_i(t)
+\sum_{j\neq i}^n \omega^{ji}_{\sigma(t)}x_{j}(t). 
\end{equation*}
Let us allow the following box constraint
\begin{equation}\label{eq:alphaConstraint}
0\leq \alpha_\ell \leq \bar \alpha_\ell
\end{equation}
on the amount of doses. Under this scenario, we consider the following optimal intervention problem:

\begin{prob}[Optimal intervention problem]\label{prb:intervention}
Let $\bar C$ be a positive constant. Assume that $\sigma$ is a semi-Markov process satisfying Assumption~\ref{asm:}.\ref{asm:sojourn}. Find the set of dose amounts~$\alpha=(\alpha_1, \dotsc, \alpha_L)$ to maximize the exponential decay rate of the system~$\Sigma'$ while satisfying 
\begin{equation}\label{eq:calpha}
C(\alpha)\leq \bar C.
\end{equation}
\end{prob}

\begin{figure}[tb]
	\centering
	\includegraphics[scale=.3]{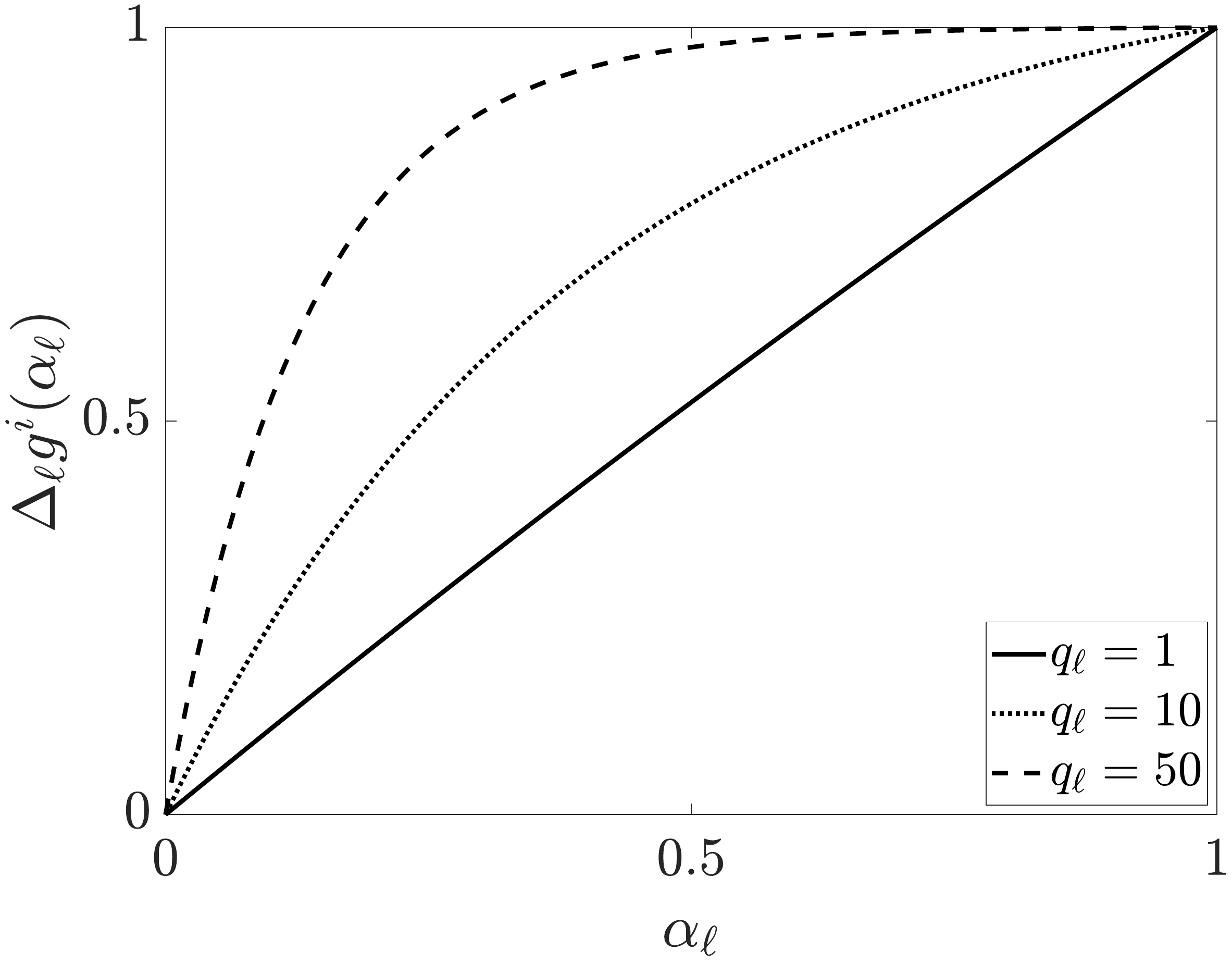}    
	\caption{Three realizations of the doasge-performance function with the parameter $\bar{\alpha}_\ell=1$, $\underline{\theta} _\ell=10$, $s_{\ell}^i=1$, and $q_\ell=1, 10, 50$.}  
	\label{fig1}                                 
\end{figure}

In this numerical example, we assume that the cost for antibiotics is linear with their dose amount, i.e., we let 
\begin{equation*}
	c_\ell(\alpha_\ell) = r_\ell\alpha_\ell
\end{equation*}
for a constant~$r_\ell > 0$ for all $\ell$. As for the suppression $\Delta_\ell g^{i}(\alpha_\ell)$ of the growth rates, we adopt the increasing function that presents diminishing marginal benefit on the dosage 
\begin{equation}\label{sim:dosage}
\Delta_\ell g^{i}(\alpha_\ell) = s_{\ell}^i\frac{\ubar \theta^{-q_\ell}_\ell-(\alpha_\ell + \ubar \theta_\ell)^{-q_\ell}}{\ubar \theta^{-q_\ell}_\ell - (\bar \alpha_\ell + \ubar \theta_\ell)^{-q_\ell}}
\end{equation}
where $\ubar \theta_{\ell}> 0$, $s_{\ell}^i \geq 0$, and $q_\ell > 0$ are parameters. These parameters allow us to realize various shapes of the suppression functions, including the dose-proportional suppression illustrated in Fig.~\ref{fig1}. We notice that the zero dose of the $\ell$th antibiotic does not change the growth rate, i.e., $\Delta_\ell g^{i}(\alpha_\ell)(0)=0$, while the maximum dose achieves the full performance $\Delta_\ell g^{i}(\bar \alpha_\ell)= s_{\ell}^i$.

Let us show that the optimal intervention problem reduces to Problem~\ref{prb:}. We introduce an auxiliary variable 
\begin{equation*}
	\theta_\ell = \ubar \theta_\ell +\alpha_\ell
\end{equation*}
that is to be optimized. If we define $\bar \theta_{\ell} = \ubar \theta_\ell + \bar \alpha_\ell$, then the constraint~\eqref{eq:alphaConstraint} is rewritten as the block constraint
\begin{equation*}
	\ubar \theta_{\ell} \leq \theta_{\ell} \leq \bar \theta_{\ell},
\end{equation*}
which can be expressed using posyonmial functions~\cite{Preciado2014}. Therefore, Assumption~\ref{asm:}.\ref{asm:posyconstraint} is satisfied. Let us define the variable~$\theta = (\theta_1, \dotsc, \theta_L)$. Then, we can rewrite the system~$\Sigma'$ into the form~\eqref{eq:sigmaTheta}, where the matrices $A_1(\theta)$, \dots, $A_N(\theta)$ are defined by
\begin{equation*}
	[A_k(\theta)]_{ii} =
	\tilde g^i_{k}+\sum_{\ell=1}^{L}{s_{\ell}^i}({\ubar \theta^{-q_\ell}_\ell - \bar \theta^{-q_\ell}_\ell})^{-1}\theta_\ell^{-q_\ell}
\end{equation*}
with
$\tilde g^i_{k} = g^i_{k}-\omega^{ii}_k-\sum_{\ell=1}^{L}{s_{\ell}^i\ubar \theta^{-q_\ell}_\ell}/{(\ubar \theta^{-q_\ell}_\ell - \bar \theta^{-q_\ell}_\ell)}$, and 
\begin{equation*}
[A_k(\theta)]_{ij} = \omega^{ij}_{k}
\end{equation*}
for $i\neq j$. Therefore, if we define the diagonal matrix~$M_k = \diag(\tilde g^1_{k}, \dotsc, \tilde g^n_{k})$, then each entry of the matrix~$A_k(\theta)-M_k$ is a posynomial in the variables~$\theta$ or zero. Hence, Assumption~\ref{asm:}.\ref{asm:metzler} is satisfied. Furthermore, the cost constraint~\eqref{eq:calpha} can be rewritten as 
\begin{equation*}
	\sum_{\ell=1}^L {r_\ell \theta_\ell} \leq \bar C + \sum_{\ell=1}^L {r_\ell\ubar \theta_\ell}
\end{equation*}
in terms of posynomials of the variable~$\theta$. Since all the conditions in Assumption~\ref{asm:} are satisfied, the optimal intervention problem can be efficiently solved by convex optimization as shown in Theorem~\ref{thm:}.

For simplicity of presentation, we focus on the case of $n=N=2$ in this numerical example. Throughout the simulation, we fix a part of the parameters as follows: $\omega^{21}_1=\omega^{12}_1=0.1$,  $\omega^{21}_2=\omega^{12}_2=0.5$, $g^1_1=1$, $g^2_1=-0.5$, $g^1_2=-1$, $g^2_2=0.5$, $\bar{\alpha}_1=\bar{\alpha}_2=1$, $\ubar{\theta}_1=\ubar{\theta}_2=10$, $q_1=q_2=0.01$, $s_{1}^1=s_{1}^2=s_{2}^1=s_{2}^2=1$, $c_1=c_2=1$, and $\bar{C}=2$. Also, we assume that the environment keeps switching from one to another, and that the sojourn time of each environment follows the Weibull distribution having the probability density function 
\begin{equation}\label{sim6}
f_{ij}(h_{ij})=
\begin{cases}
\gamma_{ij}\frac{k_{ij}}{\lambda_{ij}}\left(\frac{h_{ij}}{\lambda_{ij}}\right)^{k_{ij}-1}e^{-({h_{ij}}/{\lambda_{ij}})^{k_{ij}}}, 
\\
\quad \quad \quad\quad \quad \quad \quad \quad \quad \mbox{if $0<h_{ij} \leq T$} ,\\
0,\quad \mbox{otherwise,}
\end{cases}
\end{equation}
where $\lambda_{ij}$ is the range parameter for adjusting the expectation of the sojourn time $h_{ij}$, and $k_{ij}$ is the shape parameter. We truncate this density function at a finite time $T$ to satisfy Assumption~\ref{asm:}.\ref{asm:sojourn}, and $\gamma_{ij}$ is the constant for normalizing the integral of the truncated density function. We remark that setting $k_{ij}=1$ recovers the case of sojourn times following exponential distributions (i.e., the Markovian case). 

\begin{figure}[tb]
	\centering
	\includegraphics[scale=.3]{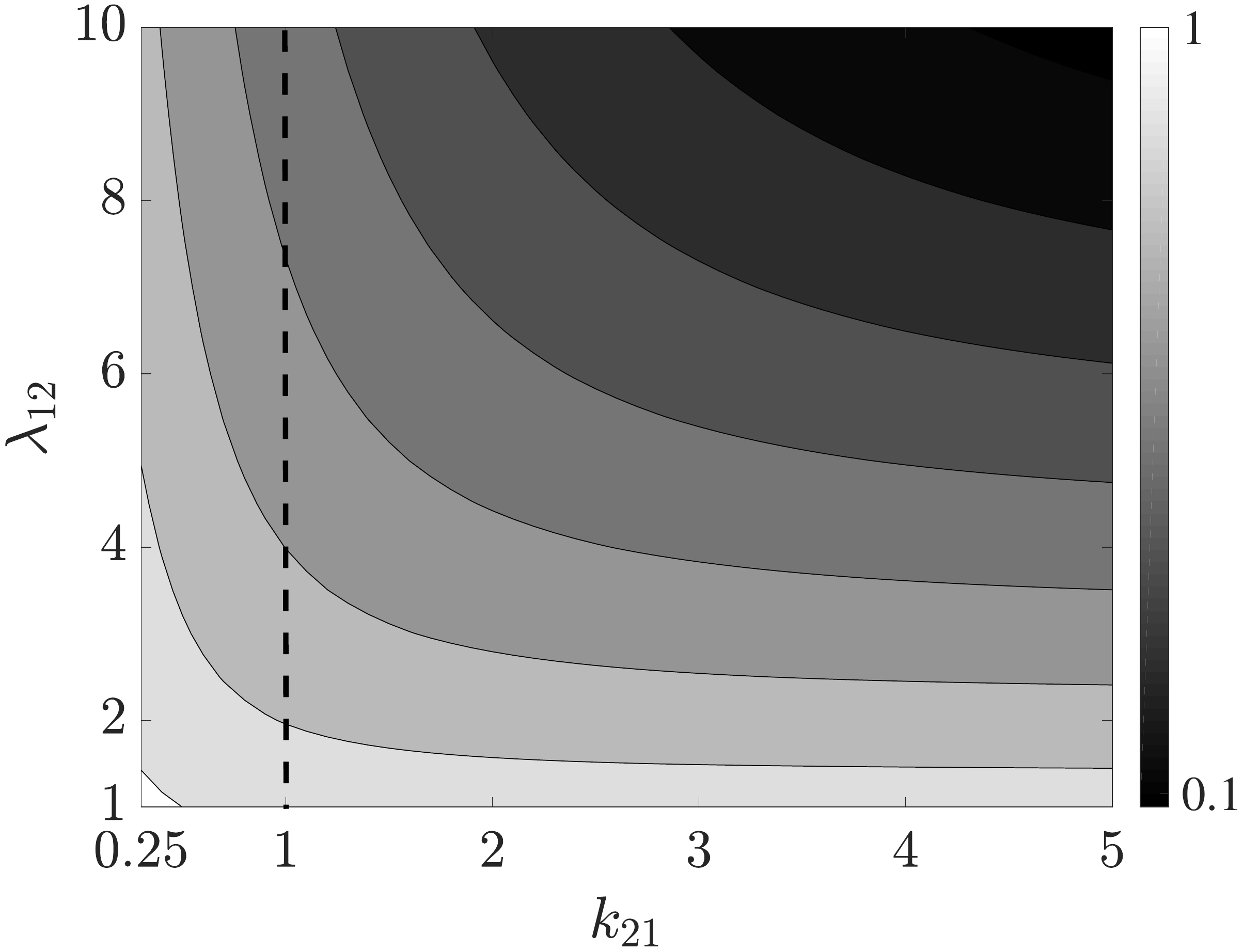}   
	\caption{{The optimized exponential decay rate} for various values of $\lambda_{12}$ and $k_{21}$. Dashed line indicates the optimized decay rate under the Markov process ($k_{12}=k_{21}=1$).}  
	\label{fig4} 
\end{figure}

 \begin{figure}[tb]
	\centering
	\includegraphics[scale=.3]{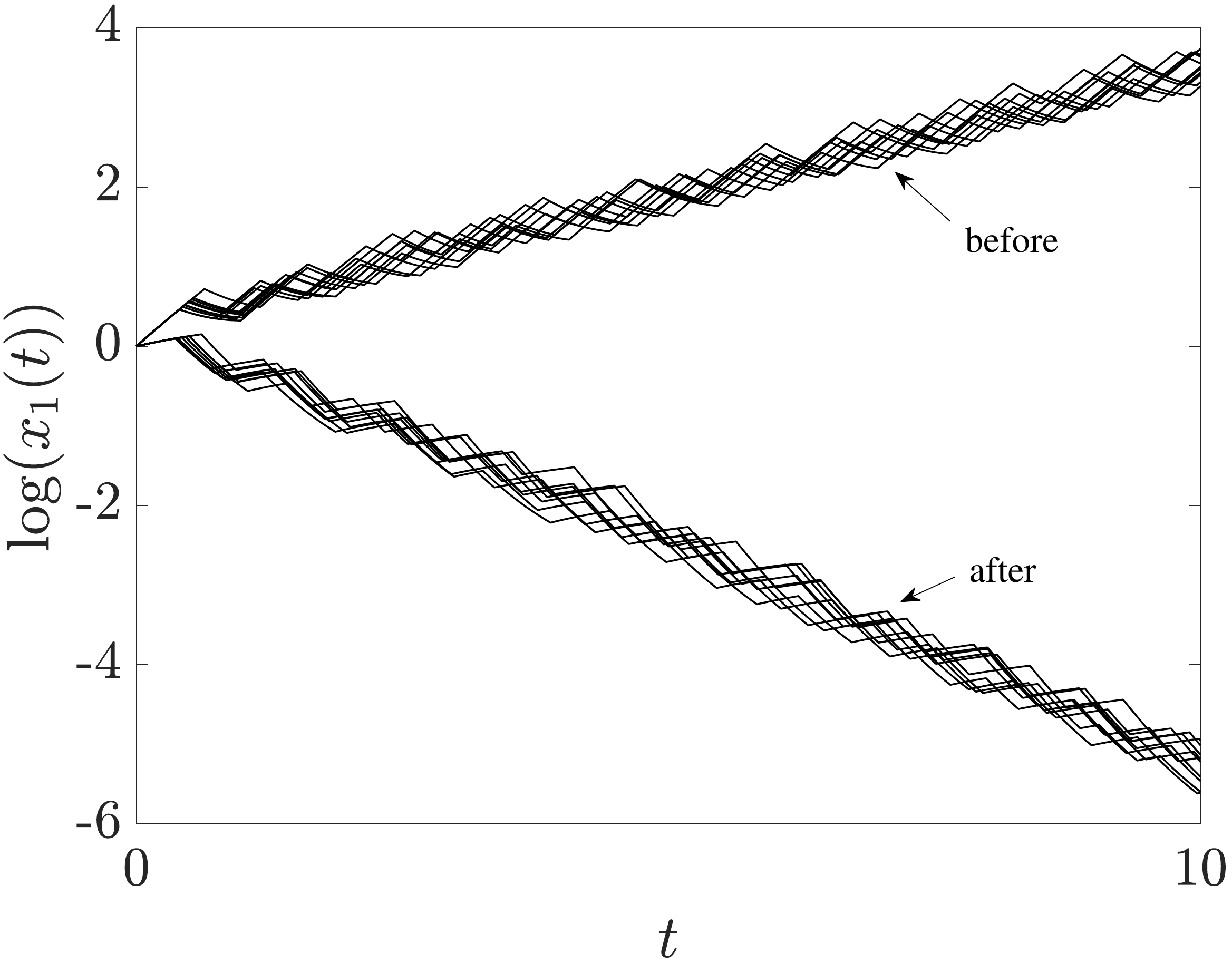}
	\hspace{0cm}   
	\caption{10 realizations of the biological populations of phenotype $1$ in the log form when $E[h_{12}]=E[h_{21}]=6$ and $\lambda_{12}=\lambda_{21}=6.67$ and $k_{12}=k_{21}=5$. In this situation, $\sigma(t)$ follows the semi-Markov process.}
	\label{fig2}     
\end{figure}

\begin{figure}[tb]
	\centering
	\includegraphics[scale=.3]{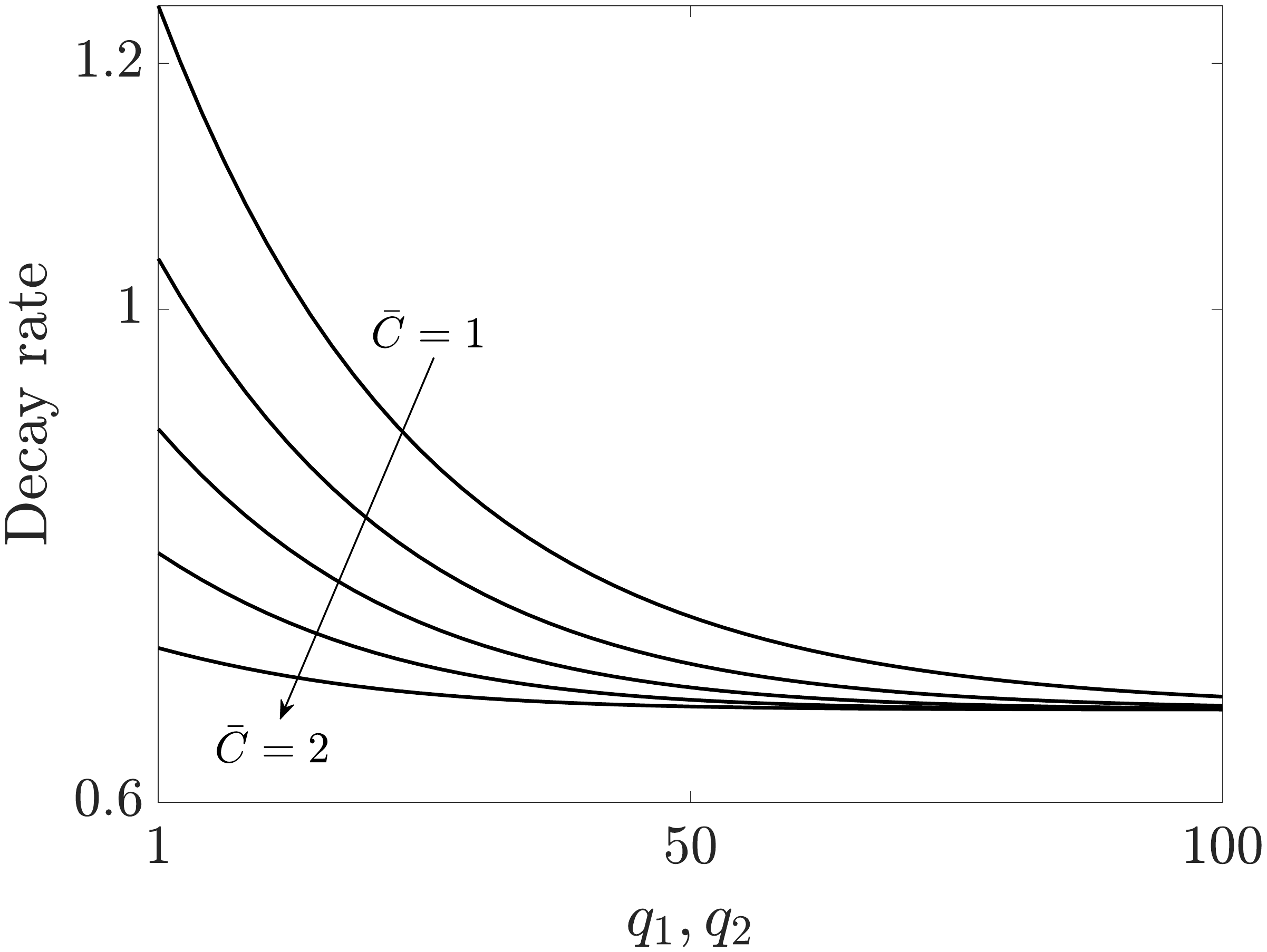}
	\hspace{0cm}
	\caption{The optimized exponential decay rate for $\bar{C}=1$, $1.2$, $1.4$, $1.6$, $1.8$, and $2$ with the variation of parameters of the dosage-performance functions $q_{1}=q_{2}$. } 
	\label{fig5}                                  
\end{figure}

In our simulation, we investigate the trade-off between the parameter $\lambda_{ij}$ and $k_{ij}$ to see the impact of the extension from Markov process to semi-Markov process. For problem solving, we adopt the commonly used off-the-shelve software for convex optimization problem: \textbf{fmincon} routine in MATLAB. For this purpose, we consider the situation in which the parameter $\lambda_{ij}$ and $k_{ij}$ of the probability density function~$f_{ij}$ in \eqref{sim6} can be tuned under the constraint of equally fixed expectation $E[h_{12}]=E[h_{21}]=6$. For various values of $\lambda_{12}$ and $k_{21}$, we present the values of the optimized exponential decay rate in Fig.~\ref{fig4}. We can observe that the optimized decay rate nontrivially depends on $k_{21}$ (i.e., the shape of the density~$f_{21}$). Specifically, when $k_{21} >1$, modeling the bet-hedging population via semi-Markov process can increase the stability of system. On the other hand, in the case of $k_{21} < 1$, we can see that the stability deteriorates. This observation shows that the trade-off between the parameter of semi-Markov process is not trivial on the stability of system. Also, it is clearly seen that our proposed framework extended the possible optimized solutions compared with the result in Markov jump linear systems, in which the optimized solutions are only represented by the dashed line in Fig.~\ref{fig4}. The biological population of phenotype 1 before and after medical intervention is illustrated in Fig.~\ref{fig2}.

Let us also investigate the dependency of the optimized decay rate on the shape parameter $q_{\ell}$ of the dosage-performance function (\ref{sim:dosage}). In this simulation, we change the values of $q_1$ and $q_2$ from $1$ to $100$ under the constraint~$q_1=q_2$. We also change the value of the budget~$\bar C$ in the interval $[1, 2]$. We solve the optimal intervention problem for various pairs of $q_1=q_2$ and $\bar C$ and obtain the optimized decay rates, as shown in Fig.~\ref{fig5}. We can see that for the fixed budget on total cost~$\bar{C}$, the increase on $q_{\ell}$ results the smaller decay rate, i.e., the stronger the diminishing property of the antibiotic, the higher decay rate can be obtained.

\section{Conclusion}

This paper studied the stabilization problem of positive semi-Markov jump linear systems. By utilizing the spectral property of nonnegative matrices, we proposed a novel computation framework that the optimal performance of the system can be formulated to a convex optimization problem which is solved by optimizing the spectral radius of the matrix under the budget-constrained of the system parameter. Then, we checked the validity through a simulation example of the biological propagation which illustrates the relations among these parameters.



\begin{thebibliography}{99}
\bibitem{Costa2013}
O. L. V. Costa, M.~D. Fragoso, and M.~G. Todorov: Continuous-time Markov Jump Linear Systems,
\textit{Springer}, 2013.

\bibitem{Bowling2006}
A.~P.~Bowling: Dynamic performance, mobility, and agility of multilegged robots,
\textit{Journal of Dynamic Systems, Measurement, and Control}, Vol.~128, No.~4, pp.~765--777, 2006.

\bibitem{Ogura2015c}
M.~Ogura and V.~M. Preciado: Stability of spreading processes over time-varying large-scale
  networks,
\textit{IEEE Transactions on Network Science and Engineering}, Vol.~3, No.~1, pp.~44--57, 2016.

\bibitem{Hespanha2007a}
J.~P. Hespanha, P.~Naghshtabrizi, and Y.~Xu: A survey of recent results in networked control systems,
\textit{Proceedings of the IEEE}, Vol.~95, No.~1, pp.~138--162, 2007.

\bibitem{Ji1990}
Y.~Ji and H.~J.~Chizeck: Controllability, stabilizability, and continuous-time Markovian jump linear quadratic control,
\textit{IEEE Transactions on Automatic Control}, Vol.~35 No.~7, pp.~777--788, 1990. 



\bibitem{Costa1998a}
O.L.V. Costa and R.P. Marques: Mixed $H_2$/$H_\infty$-control of discrete-time Markovian jump linear systems,
\textit{IEEE Transactions on Automatic Control}, Vol.~43, No.~1, pp.~95--100, 1998. 

\bibitem{Geromel2015}
J.~C. Geromel and G.~W. Gabriel: Optimal $\mathcal H_2$ state feedback sampled-data control design of Markov Jump Linear Systems,
\textit{Automatica}, Vol.~54, pp.~182--188, 2015.

\bibitem{Mukaidani2017}
H.~Mukaidani, H.~Xu, T.~Shima, and V.~Dragan: A stochastic multiple-leader–follower incentive stackelberg strategy for markov jump linear systems,
\textit{IEEE Control Systems Letters}, Vol.~1, No.~2, pp.~250--255, 2017.

\bibitem{Ogura2012b}
M.~Ogura and C.~F. Martin: Generalized joint spectral radius and stability of switching systems,
\textit{Linear Algebra and its Applications}, Vol.~439, No.~8, pp.~2222--2239, 2013. 

\bibitem{Hosoe2019}
Y.~Hosoe and T.~Hagiwara: Equivalent stability notions, Lyapunov inequality, and its
  application in discrete-time linear systems with stochastic dynamics
  determined by an i.i.d. process,
\textit{IEEE Transactions on Automatic Control}, Vol.~64, No.~11, pp.~4764--4771, 2019.

\bibitem{Johnson1989}
B.~W.~Johnson: Design and Analysis of Fault-Tolerant Digital Systems,
\textit{Addison-Wesley}, 1989.

\bibitem{Ogura2013f}
M.~Ogura and C.~F. Martin: Stability analysis of positive semi-Markovian jump linear systems
  with state resets,
\textit{SIAM Journal on Control and Optimization}, Vol.~52, pp.~1809--1831, 2014.

\bibitem{Schioler2014}
H.~Schioler, M.~Simonsen, and J.~Leth: Stochastic stability of systems with semi-Markovian switching,
\textit{Automatica}, Vol.~50, No.~11, pp.~2961--2964, 2014.

\bibitem{Huang2012}
J.~Huang and Y.~Shi: Stochastic stability and robust stabilization of semi-Markov jump
  linear systems,
\textit{International Journal of Robust and Nonlinear Control}, Vol.~23, No.~18, pp.~2028--2043, 2012.

\bibitem{Chen2016b}
L.~Chen, X.~Huang, and S.~Fu: Observer-based sensor fault-tolerant control for semi-Markovian jump
  systems,
\textit{Nonlinear Analysis: Hybrid Systems}, Vol.~22, pp.~161--177, 2016.

\bibitem{Jiang2018}
B.~Jiang, Y.~Kao, H.~R.~Karimi, and C.~Gao: Stability and stabilization for singular switching semi-Markovian
  jump systems with generally uncertain transition rates,
\textit{IEEE Transactions on Automatic Control}, Vol.~63, No.~11, pp.~3919--3926, 2018.

\bibitem{Shen2018}
H.~Shen, F.~Li, S.~Xu, and V.~Sreeram: Slow state variables feedback stabilization for semi-Markov jump
  systems with singular perturbations,
\textit{IEEE Transactions on Automatic Control}, Vol.~63, No.~8, pp.~2709--2714, 2018.

\bibitem{Wei2017}
Y.~Wei, J.~H. Park, J.~Qiu, L.~Wu, and H.~Y. Jung: Sliding mode control for semi-Markovian jump systems via output
  feedback,
\textit{Automatica}, Vol.~81, pp.~133--141, 2017.

\bibitem{Boyd2007}
S.~Boyd, S.~J.~Kim, L.~Vandenberghe, and A.~Hassibi: A tutorial on geometric programming,
\textit{Optimization and Engineering}, Vol.~8, No.~1, pp.~67--127, 2007.



\bibitem{kingman1961}
J.~F.~C. Kingman: A convexity property of positive matrices,
\textit{Quarterly Journal of Mathematics}, Vol.~12, No.~1, pp.~283--284, 1961.

\bibitem{Kussell2005}
E.~Kussell and S.~Leibler: Phenotypic diversity, population growth, and information in
  fluctuating environments,
\textit{Science}, Vol.~309, pp.~2075--2078, 2005.

\bibitem{Farina2000}
L.~Farina and S.~Rinaldi: Positive Linear Systems: Theory and Applications,
\textit{Wiley-Interscience}, 2000.

\bibitem{Ebihara2017a}
Y.~Ebihara: $H_2$ analysis of LTI systems via conversion to externally positive
  systems,
\textit{IEEE Transactions on Automatic Control}, Vol.~63, No.~8, pp.~2566--2572, 2018.

\bibitem{Janssen2006}
J.~Janssen and R.~Manca: Applied Semi-Markov Processes,
\textit{Springer-Verlag}, 2006.

\bibitem{Feng1992}
X.~ Feng, K.~A.~Loparo, Y.~Ji, and H.~J.~Chizeck: Stochastic stability properties of jump linear systems,
\textit{IEEE Transactions on Automatic Control}, Vol.~37, pp.~38--53, 1992.

\bibitem{Preciado2014}
V.~M.~Preciado, M.~Zargham, C.~Enyioha, A.~Jadbabaie, and
  G.~J. Pappas: Optimal resource allocation for network protection against spreading
  processes,
\textit{IEEE Transactions on Control of Network Systems}, Vol.~1, No.~1, pp.~99--108, 2014.

\bibitem{Rantzer2018}
A.~Rantzer and M.~E.~Valcher: A tutorial on positive systems and large scale control,
\textit{57th IEEE Conference on Decision and Control}, No.~978, pp.~3686--3697, 2018.

\bibitem{Ogura2019c}
M.~Ogura, M.~Kishida, and J.~Lam: Geometric programming for optimal positive linear systems,
\textit{IEEE Transactions on Automatic Control} \upshape (accepted for
  publication), 2020.

%
%
%
%
%












\bibitem{Cohen1981}
J.~E. Cohen: Convexity of the dominant eigenvalue of an essentially nonnegative
  matrix,
\textit{Proceedings of the American Mathematical Society}, Vol.~81, No.~4, pp.~657--658, 1981.

\end{thebibliography}

\begin{biography}
\profile{n}{Chengyan Zhao}{
He received his B.S. and M.S. from Northeastern University, China, in 2011 and 2013, respectively. Now, he is a Ph.D. student in the Department of Science and Technology, Nara Institute of Science and Technology, Japan. His research interests include positive systems, switched linear systems and complex networks control. He is a member of IEEE.
}  

\profile{m}{Masaki Ogura}{Masaki Ogura is an Associate Professor in the Graduate School of Information Science and Technology at Osaka University, Japan. Prior to joining Osaka University, he was a Postdoctoral Researcher at the University of Pennsylvania, USA and an Assistant Professor at the Nara Institute of Science and Technology, Japan. His research interests include network science, dynamical systems, and stochastic processes with applications in networked epidemiology, design engineering, and biological physics. He was a runner-up of the 2019 Best Paper Award by the IEEE Transactions on Network Science and Engineering and a recipient of the 2012 SICE Best Paper Award. He is an Associate Editor of the Journal of the Franklin Institute.
}  

\profile{f}{Kenji Sugimoto}{ Kenji Sugimoto is a Professor at the Graduate School
of Science and Technology, Nara Institute of Science
and Technology, Japan. After receiving Master's degree
from Kyoto University, he worked for Mitsubishi Electric
Corporation. He was an Assistant Professor in Kyoto
University, Associate Professors in Okayama and Nagoya
Universities, before the current position. His research
interest includes control theory and its application.
He is a member of IEEE and ISCIE, and an Editor of
SICE Japanese Transactions. He has a doctor's degree
of engineering.
}  

\end{biography}

\end{document}